\documentclass[12pt]{iopart}

\def\sNN{\mbox{$\sqrt{s_{_{NN}}}$}}

\newcommand{ \vtwoEcc} {v_2/\varepsilon}

\newcommand{ \mean }[1]{\left\langle #1 \right\rangle} 
\usepackage{graphicx} 
\begin{document}

\title[Parton Collectivity]{Parton Collectivity from {\sc RHIC} to the {\sc LHC}}

\author{R. Snellings}

\address{Nikhef}
\ead{Raimond.Snellings@nikhef.nl}
\begin{abstract}
Anisotropic flow is recognized as one of the main observables providing information on the
early dynamics in heavy-ion collisions. The large elliptic flow observed at {\sc RHIC} is considered to be evidence 
for almost perfect liquid behavior of the strongly coupled Quark Gluon Plasma produced in the collisions. 
In this report we review our current understanding of this new state of matter and investigate the predictions for
anisotropic flow at the {\sc LHC}. 
\end{abstract}

\vspace{2pc}

\section{Introduction}

Relativistic heavy-ion collisions are a unique
tool to study the Equation of State ({\sc EoS}) of extremely hot and dense matter 
under controlled conditions.
The system of hot and dense matter created in a 
heavy-ion collision will expand and cool down. 
During this expansion the system probes a range of energy densities and temperatures, 
and possibly different phases.
The collective dynamics of the system is revealed in the particle yield as function of transverse momentum,
which is characterized by the temperature and transverse flow velocity of the system at freeze-out.
Because heavy-ions are not point-like, 
the created system in non-central collisions has an
azimuthal anisotropy in coordinate space which translates, due to multiple
interactions, into an azimuthal anisotropy in momentum space.
The elliptic flow $v_2$ is defined as the second coefficient of a Fourier expansion of this
azimuthal anisotropy~\cite{Voloshin:1994mz}. This elliptic flow is particularly 
sensitive to the early dynamics of the collision, for a recent review see~\cite{Voloshin:2008dg}.

\begin{figure}[thb]
    \begin{center}
    \includegraphics[width=0.9\textwidth]{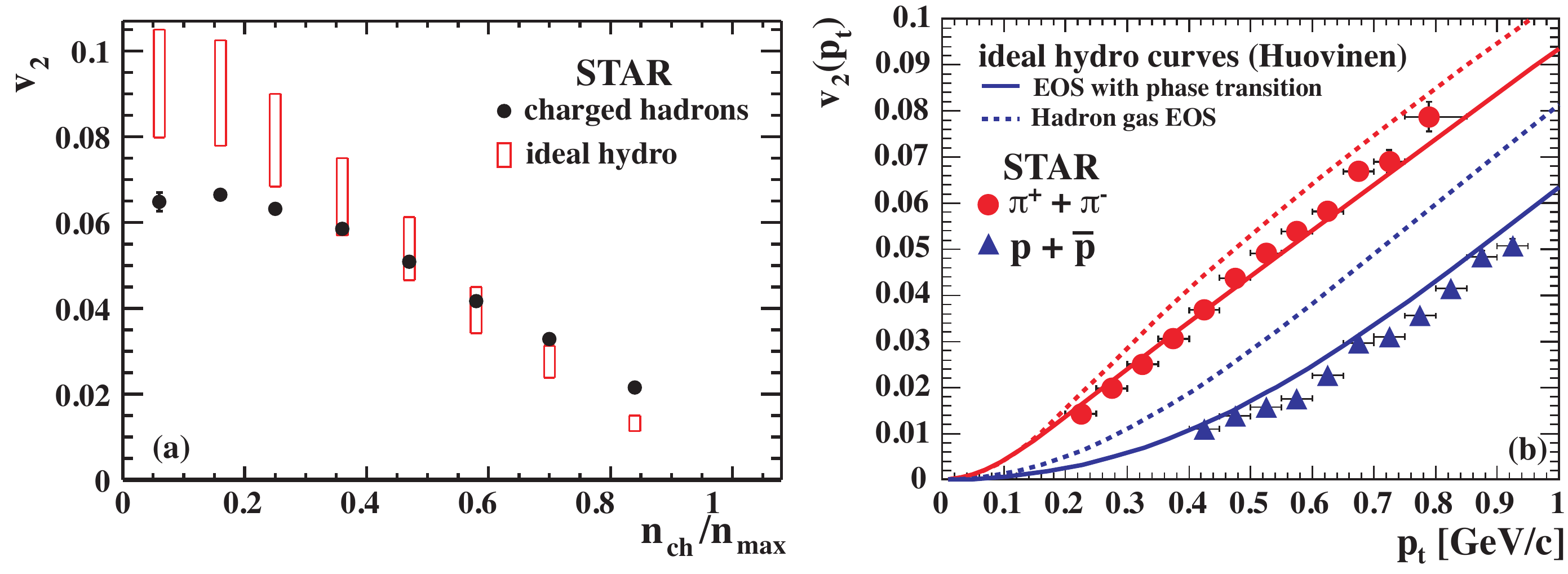}
      \caption{ (Color on-line) 
     	(a) Elliptic flow (solid points) as a function of centrality
  	 $n_{ch}/n_{max}$. 
	The open rectangles show a range of
 	 values expected for $v_2$ in the ideal hydrodynamic limit~\cite{Ackermann:2000tr}.
	(b) Elliptic flow of pions and protons as function of transverse 
	momentum~\cite{Adler:2001nb}.
	The curves are hydrodynamical model calculations using a hadron gas {\sc EoS} (dashed curve) 
	and an {\sc EoS} which incorporates the {\sc QCD} phase transition (full curve).
      \label{elliptic_flow} }
    \end{center}
\end{figure}
Generally speaking, large values of elliptic flow are considered to be 
signs of hydrodynamic behavior as was first put forward by
Ollitrault~\cite{Ollitrault:1992bk}.
In ideal hydrodynamics $v_2$ is proportional to the spatial eccentricity 
with a magnitude which depends on the {\sc EoS}.
The spatial eccentricity is defined by
\[
\varepsilon = \frac{\langle y^2 - x^2 \rangle}{\langle y^2 + x^2 \rangle}
\]
where $x$ and $y$ are the spatial coordinates of the colliding nucleons in the plane
perpendicular to the collision axis and where the brackets 
denote an average.
In practice $\varepsilon$ is not a measured quantity but obtained from Glauber or Color Glass
Condensate ({\sc CGC}) model calculations.
The successful description of the measured elliptic flow~\cite{Ackermann:2000tr,Adler:2001nb} 
at {\sc RHIC} by ideal hydrodynamical models, as shown in Fig.~\ref{elliptic_flow}, lead to the 
conclusion that the system created at RHIC behaves like a perfect liquid~\cite{perfectLiquid}. 

\section{The perfect liquid?}

\begin{figure}[htb]
    \begin{center}
    \includegraphics[width=0.45\textwidth]{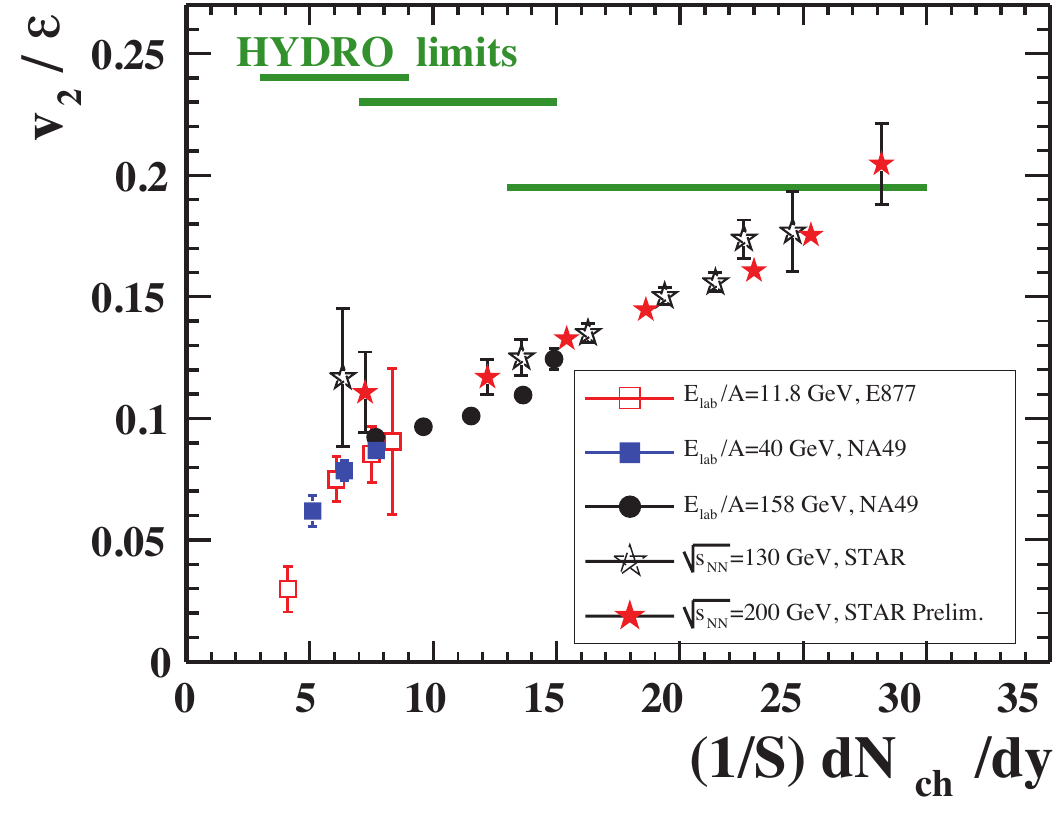}
      \caption{ (Color on-line) 
     	Compilation of $v_2/\varepsilon$ data~\cite{Alt:2003ab,Adler:2002pu} versus particle density at midrapidity. 
	Green lines indicate ideal hydrodynamic predictions for AGS, SPS and RHIC collisions energies
      \label{v2eps} }
    \end{center}
\end{figure}

Also from Fig.~\ref{v2eps} 
it is seen that the measured $v_2$  
scaled by the spatial eccentricity versus particle density reaches the expected ideal 
hydrodynamic values but only  for the more central collisions at the highest RHIC center of mass 
energy~\cite{Alt:2003ab,Adler:2002pu}. 
Discrepancies are observed for peripheral collisions, lower energies, and regions away from mid-rapidity 
which indicates that here the elliptic flow has significant non-ideal hydro contributions. 

Much of this discrepancy can be explained by incorporating the viscous contribution from the 
hadronic phase~\cite{Teaney:2001av,Teaney:2000cw,Hirano:2005wx,Hirano:2005xf}.
However, we expect that also the hot and dense phase must deviate from an 
ideal hydrodynamic description. 
Kovtun, Son and Starinets ({\sc KSS})~\cite{Kovtun:2004de}, showed that conformal field theories 
with gravity duals have a ratio of shear viscosity $\eta$ to entropy density $s$ of,  in natural units, $\eta/s = 1/4\pi$.
They conjectured that this value is a lower bound for any relativistic thermal field theory. 
In addition, Teaney~\cite{Teaney:2003kp} pointed out that very small shear viscosities, 
of the magnitude of the bound, would already lead to a significant reduction in the predicted elliptic flow.

\begin{figure}[htb]  
\begin{center}
  \includegraphics[width=1.\textwidth]{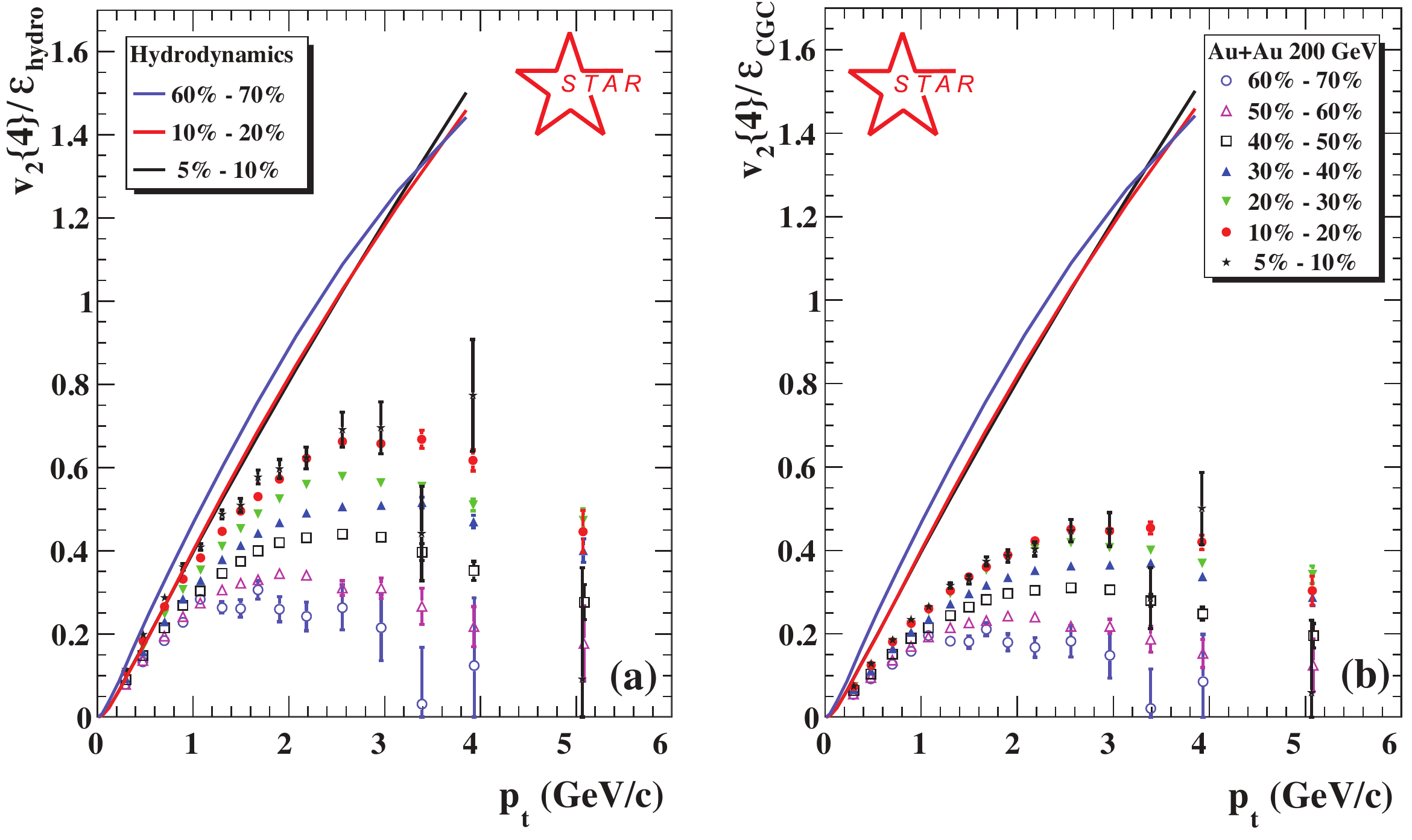}
\end{center}
\caption{(color online)
Scaled elliptic flow $\vtwoEcc$ versus transverse momentum for eccentricities used in ideal hydrodynamics (left) 
and CGC (right) compared to ideal hydrodynamics~\cite{YutingThesis,Tang:2008if}.}
\label{fig:v2ecc_pt}
\end{figure}

Teaney~\cite{Teaney:2003kp} also pointed out that 
the reduction of elliptic flow due to non-zero $\eta/s$ would become larger at higher transverse momenta.
In Fig.~\ref{fig:v2ecc_pt} 
ideal hydro calculations are compared to 
{\sc STAR} measurements of $v_2/\varepsilon$ as function of transverse momentum 
for different centralities. 
In the figure the initial eccentricity is calculated by either a Glauber model (left panel) or 
a CGC model~\cite{Drescher:2007cd} (right panel).
The figures show that with {\sc CGC} initial eccentricities the magnitude of $\vtwoEcc$ is lower than
with Glauber initial eccentricities~\cite{Hirano:2005xf}. 

The ideal hydro curves in Fig.~\ref{fig:v2ecc_pt} correspond to calculations~\cite{HuovinenPersonal} 
using an initial eccentricity from a Glauber model. 
As was shown already in Fig.~\ref{v2eps} the elliptic flow in the more central collisions reach the ideal hydrodynamical
values, provided that Glauber initial conditions are used.
It is seen in the right panel of Fig.~\ref{fig:v2ecc_pt} that this is not true anymore when {\sc CGC} eccentricites are used 
which clearly illustrates that constraints on the initial conditions are crucial.
Nevertheless, for both Glauber and CGC initial eccentricities, it is seen that 
going from peripheral to central collisions the ratio $v_2/\varepsilon$ increases and that the transverse momentum where 
$v_2$ is maximal rises. 
Both these observations are consistent with a decrease of $\eta/s$ from peripheral to central 
collisions~\cite{Teaney:2003kp}. 

One approach to quantify the contribution from $\eta/s$ is to incorporate 
viscous corrections into the hydrodynamic model. 
A drawback of this approach is that solving viscous relativistic hydrodynamical equations is 
not straight forward and that the results are sensitive to not only $\eta/s$ but as in ideal hydrodynamics also to, 
among others, the initial spatial eccentricity~\cite{Hirano:2005xf} and the {\sc EoS}~\cite{Huovinen:2005gy}.
Nevertheless, tremendous progress has recently been made in this 
area~\cite{Heinz:2005bw,Baier:2006um,Chaudhuri:2006jd,Romatschke:2007mq,Dusling:2007gi,Song:2008si,
Luzum:2008cw,Song:2007fn,Muronga:2006zw,Huovinen:2008te,Betz:2008me}. 

\begin{figure}[htb]
    \begin{center}
    \includegraphics[width=1.\textwidth]{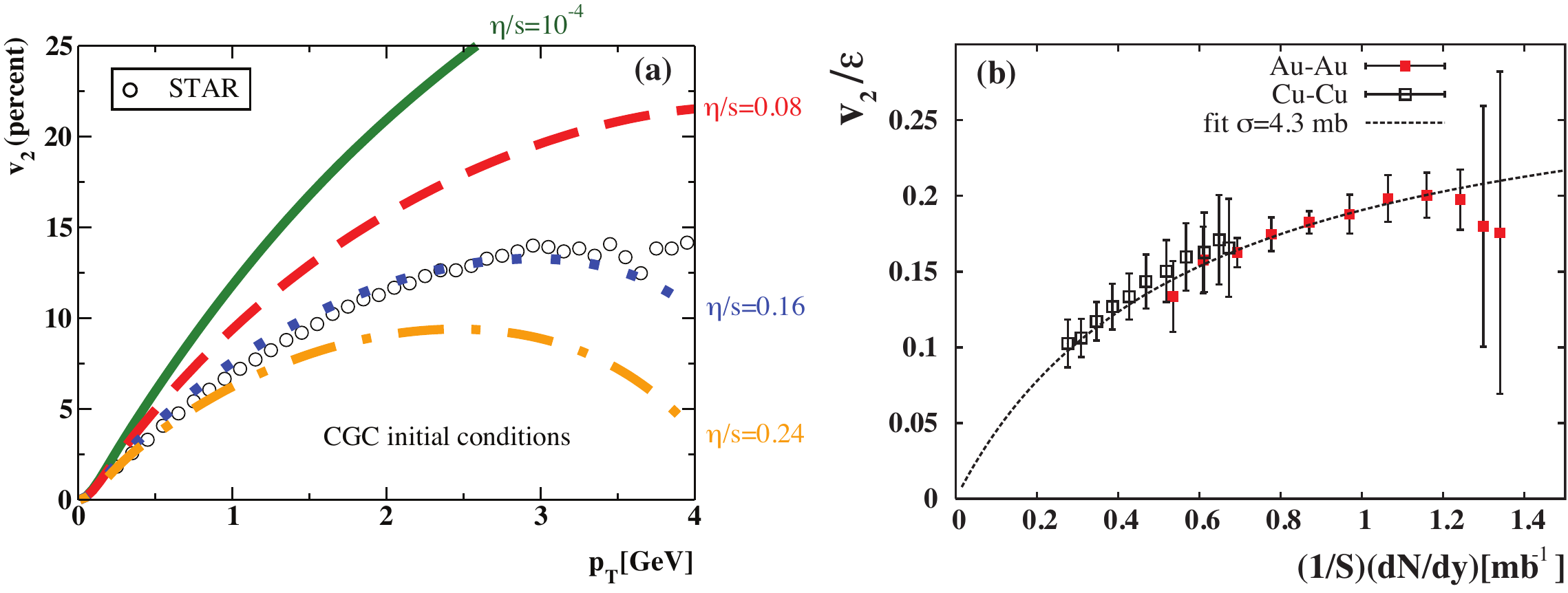}
      \caption{ (Color on-line) 
         (a) Elliptic flow versus transverse momentum using CGC initial conditions in hydrodynamics with viscous corrections.
         The {\sc STAR} measurements (open circles) have been approximately corrected for fluctuations and nonflow.
         (b) Fit to $v_2/\varepsilon$ versus particle density in terms of Knudsen number for Cu+Cu and 
	Au+Au~\cite{Drescher:2007cd}. 
      \label{v2viscous} }
    \end{center}
\end{figure}
Figure~\ref{v2viscous}a shows, as an example, the measured transverse momentum dependence of the 
elliptic flow for charged particles compared to a viscous hydro model calculation~\cite{Luzum:2008cw}. 
Here, the initial eccentricity is 
based on a {\sc CGC} model calculation and results are shown for various values of $\eta/s$. 
It is again clear from Fig.~\ref{v2viscous} that the difference between ideal hydrodynamics (full curve) and the measured flow
is substantial and that this difference increases with transverse momentum.
Good agreement with the data is obtained when $\eta/s$ is about twice the {\sc KSS} bound of 0.08.

Motivated by transport calculations, another promising approach to quantify the possible discrepancy 
with ideal hydrodynamics is a description of $v_2/\varepsilon$ versus particle density 
in terms of the {\em Knudsen number}~\cite{Drescher:2007cd,Bhalerao:2005mm}. 
This Knudsen number is the mean free path of the constituents divided by the system size, and is defined
by~\cite{Drescher:2007cd,Bhalerao:2005mm}:
\[
K = \left(\sigma c_s \frac{1}{S}\frac{dN}{dy}\right)^{-1}.
\]
Here $\sigma$ is the constituent transport cross section, $c_s$ the velocity of sound and
$S$ the transverse area $4 \pi \sqrt{\mean{x^2}\mean{y^2}}$.

The scaled elliptic flow is parameterized as~\cite{Drescher:2007cd,Bhalerao:2005mm}:
\begin{equation}
  \frac{v_{2}}{\epsilon} =  \frac{h}{1+K/K_0}, 
\label{eq:knudsen}
\end{equation}
where $h$ is the ideal hydrodynamic limit of $v_{2}/\varepsilon$ and $K_0 = 0.7$ is a constant  
determined from transport calculations~\cite{Drescher:2007cd}. 
This parameterization yields correctly the hydro limit for $K \rightarrow 0$ and the 
low density limit~\cite{Heiselberg:1998es} for $K \rightarrow \infty$:
\begin{eqnarray}
\left[  \frac{v_{2}}{\epsilon} \right]_{K \rightarrow  0} &=& h - K + K^2 - ......  \nonumber \\ 
\left[  \frac{v_{2}}{\epsilon} \right]_{K \rightarrow  \infty} & \propto &  \sigma c_s \frac{1}{S}\frac{dN}{dy} \nonumber .
\label{eqv2densitylimits}
\end{eqnarray}
       
Figure~\ref{v2viscous}b shows  $v_2/\varepsilon$ measured by {\sc PHOBOS} in Cu$+$Cu  and
Au$+$Au collisions at \sNN = 200 GeV, using Glauber initial conditions for $\varepsilon$.
The curve in the figure shows a fit of Eq.~\ref{eq:knudsen} to the data with $h$ and $\sigma$ as free parameters.
Input to the fit are the quantities $dN/dy$, $\varepsilon$ and $S$ obtained from measurement and model calculations.
Clearly the parameterization describes the data well over the whole range of particle densities
and shows no sign of saturation.
Therefore the authors of the fit conclude in~\cite{Drescher:2007cd} that even for the most central collisions at the
highest {\sc RHIC} energy the deviations from ideal hydrodynamics are as large as 30\%. 
The magnitude of $\eta/s$ is obtained from the fitted value of $\sigma$ combined with an additional estimate 
of the temperature where $v_2$ develops.
The value of $\eta/s$ is found to be a few times the {\sc KSS} bound, 
which is roughly in agreement with estimates from viscous hydrodynamical model calculations.

\section{Collective flow at the LHC}

Most theoretical models~\cite{Abreu:2007kv} 
predict an increase of $v_2$ at the {\sc LHC} compared to {\sc RHIC}.
In the following we will describe estimates based on (ideal) hydrodynamics, the low density limit and the
Knudsen parameterization.
 
\begin{figure}[htb]  
\begin{center}
  \includegraphics[width=0.5\textwidth]{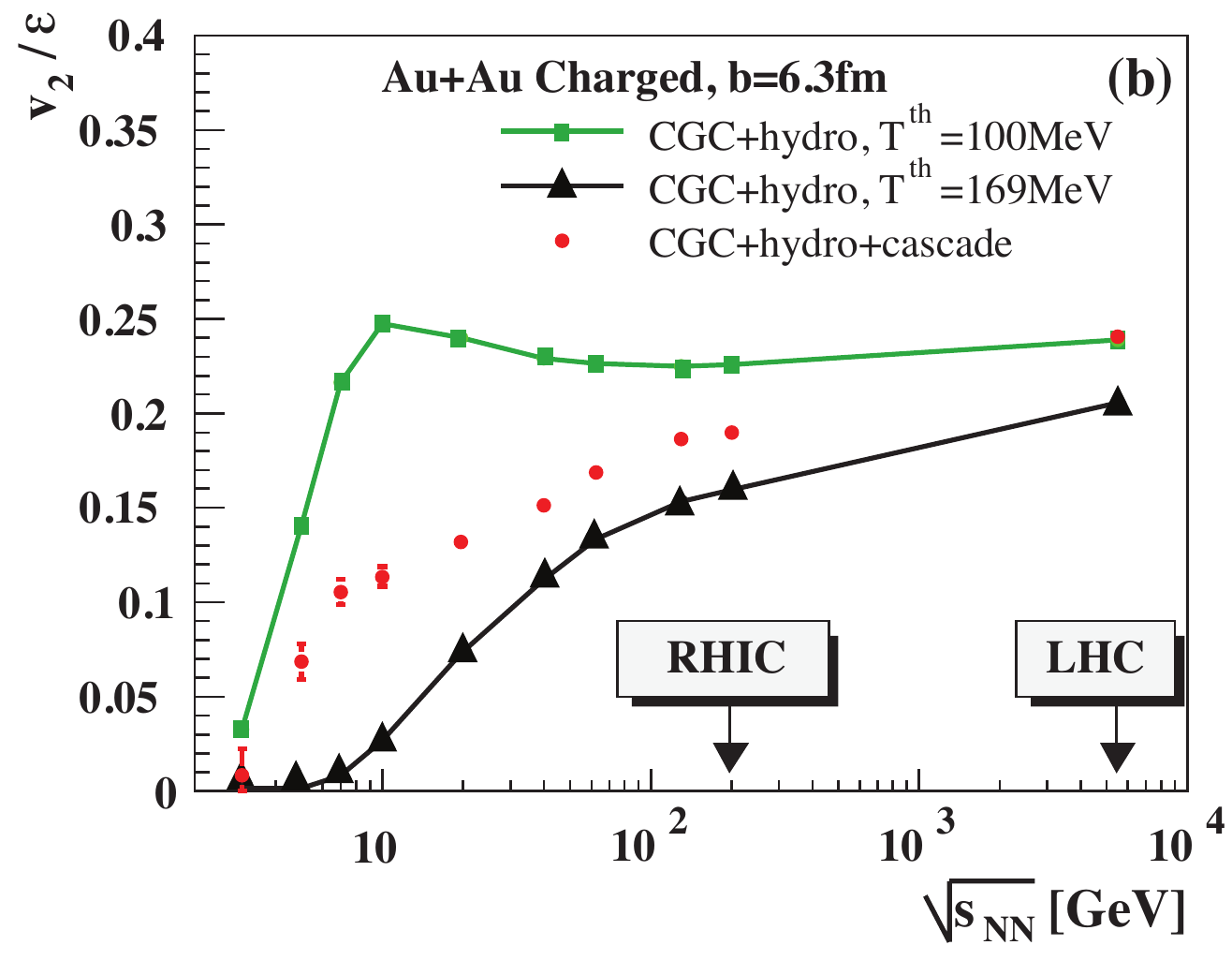}
\end{center}
\caption{(color online) $v_2$ versus center of mass energy from a ideal hydro + hadron cascade 
	model~\cite{Hirano:2007gc}.
}
\label{fig:v2hydrovis}
\end{figure}

Figure~\ref{fig:v2hydrovis} 
shows predictions of $v_2/\varepsilon$ versus center of mass energy
from a model that combines ideal hydrodynamics and a hadron cascade~\cite{Hirano:2007gc}. 
The lower curve (full triangles) shows the elliptic flow from the partonic phase, assuming ideal hydrodynamics.
The upper curve (full squares) shows this elliptic flow with the contribution from the hadronic phase 
added, here the hadronic contribution also is calculated using ideal hydrodynamics.
When using a hadron cascade instead of hydrodynamics to describe the evolution of the hadronic phase the predictions 
are obtained as shown by the full circles in Fig.~\ref{fig:v2hydrovis}. It turns out that this combination gives the best
description of the data (not shown).
It is clear from these calculations that the contribution from the partonic phase is considerably larger at the LHC than at
RHIC energies.  It is also seen that the prediction from a full ideal hydrodynamic evolution yields the same answer 
as a hydro + hadron cascade evolution at LHC energies. This indicates that the viscous contribution from the hadronic
phase is relatively unimportant at the LHC.

\begin{figure}[htb]  
\begin{center}
  \includegraphics[width=1.\textwidth]{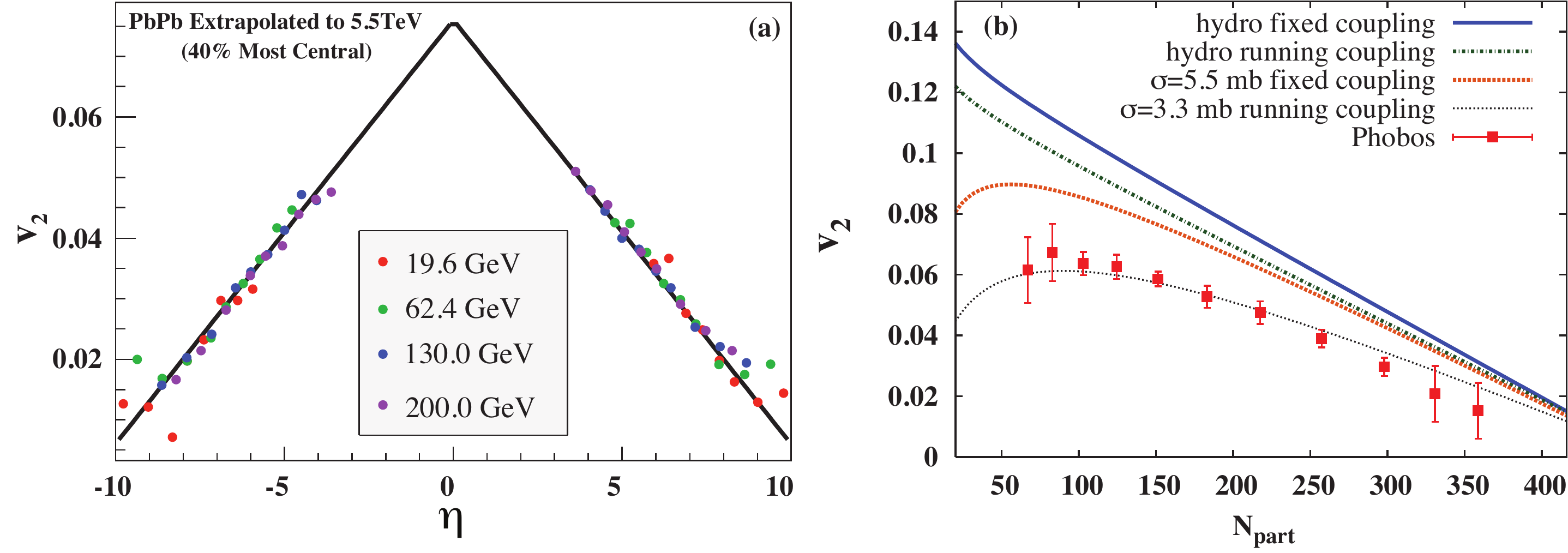}
\end{center}
\caption{(color online)
(a) $v_2$ versus $\eta$ measured by {\sc PHOBOS} at {\sc RHIC}, using longitudinal scaling to 
make a prediction for the {\sc LHC}~\cite{Busza:2007ke}.
 (b) Elliptic flow $v_2$ as function of centrality $N_{part}$ at mid-rapidity for Pb-Pb collisions 
 at the LHC~\cite{Drescher:2007uh}. 
}
\label{fig:v2LHCpredict}
\end{figure}        

In the low density limit elliptic flow is proportional to the particle multiplicity. 
It has been shown by 
PHOBOS that particle multiplicity as function of center of mass energy is well described 
using longitudinal scaling~\cite{Back:2002wb}.
It is therefore expected, in the low density limit, that also $v_2$  
exhibits longitudinal scaling, provided that the eccentricity is kept constant. 
The longitudinal scaling is demonstrated in Fig.~\ref{fig:v2LHCpredict}a where measurements of $v_2$ at
various RHIC center of mass energies are plotted versus the pseudorapity $\eta$ at the LHC.
Linear extrapolation to mid-rapidity yields an increase of about 50\% in the magnitude of $v_2$ compared to RHIC.
This increase in $v_2$ is larger than predicted by most other models.

The third prediction presented here is based on the parameterization in terms of 
the Knudsen number  (Eq.~\ref{eq:knudsen}).
The two lower curves in Fig.~\ref{fig:v2LHCpredict}b show the centrality dependence of $v_2$ 
assuming either a fixed coupling or running coupling evolution of the saturation scale $Q_s$.
The corresponding transport cross sections are $\sigma = 5.5$ and $3.3$~mb respectively.
The first value corresponds to RHIC assuming CGC initial conditions.
The upper two curves are prediction based on "ideal hydrodynamics" (assuming $v_2/\varepsilon = h$).
For fixed coupling it is seen that the Knudsen parameterization closely approaches the ideal hydro curves  
except for the most peripheral collisions.
The ideal hydrodynamical limit $h$ is a free parameter therefore the predicted magnitude
of $v_2$ itself is not fixed (however all four curves move together changing $h$). 
For the figure  the previously found {\sc RHIC} value of $h= 0.22$ is used.
The centrality dependence of $v_2$ does not depend on $h$ and, as can be seen in the figure,   
the measurement of this centrality dependence at the {\sc LHC} 
will allow us to distinguish between ideal hydro and the Knudsen parameterization.

\section{Summary}
At {\sc RHIC} the observed large elliptic flow provides compelling evidence for strongly interacting matter
which appears to behave like an almost perfect fluid. 
To quantify this the significant effects of the viscous corrections need to be
calculated. 
At  \sNN = 200 GeV, based on different model assumptions, calculations of $\eta/s$ have been performed and  
are found to be in approximate agreement, ranging for the more central collisions 
from 1--4 times the {\sc KSS} bound. 
Even though this is encouraging, it is important to realize that these 
results are all obtained based on a small set of initial conditions. 
The estimated range of $\eta/s$ values would certainly be affected if
the initial conditions were to be sufficiently different, e.g. due to strong initial flow fields.

To quantify $\eta/s$ of the partonic fluid requires knowledge of the relative contributions from the 
partonic and hadronic phase.
Detailed comparison of identified particle elliptic flow, in particular for particles which have different 
hadronic cross sections, 
should allow us to constrain this experimentally. 
In addition, comparing the center of mass energy dependence of $v_2$ in which  
the relative contribution from the partonic and hadronic phase to $v_2$ varies should 
provide additional constraints.
Elliptic flow measurements at the {\sc LHC} will provide an important contribution to this and 
in addition a decisive test which of the currently successful model descriptions is more appropriate. 

\section*{Acknowledgments}
The author would like to thank Michiel~Botje, Adrian~Dumitru, Tetsufumi Hirano, Pasi~Huovinen, Jean-Yves~Ollitrault, Art~Poskanzer,  
Aihong~Tang, Nu~Xu and Sergei~Voloshin for their contributions.
This work is supported by NWO and FOM.

\end{document}